\documentclass[reprint,prl,twocolumn,showpacs,superscriptaddress,aps,longbibliography]{revtex4-2}

\usepackage{natbib}
\usepackage[colorlinks=true,citecolor=blue]{hyperref} 
\usepackage{graphicx}
\usepackage{bm}
\usepackage{amsmath}
\usepackage{amssymb}
\usepackage[normalem]{ulem}
\usepackage{physics}
\usepackage[T1]{fontenc}
\usepackage[utf8]{inputenc}

\DeclareGraphicsExtensions{.png,.jpg,.eps}
\usepackage{xcolor}

\usepackage{stmaryrd}

\begin{document}
\title{Fingerprinting superconductors by disentangling Andreev and quasiparticle currents across tunable tunnel junctions}

\author{Petro Maksymovych}
\affiliation{Department of Materials Science and Engineering, Clemson University, Clemson, SC, USA}
\author{Sang Yong Song}
\affiliation{Materials Science and Technology Division, Oak Ridge National Laboratory, Oak Ridge, TN, USA}
\author{Benjamin Lawrie}
\affiliation{Materials Science and Technology Division, Oak Ridge National Laboratory, Oak Ridge, TN, USA}
\author{Wonhee Ko}
\affiliation{Department of Physics and Astronomy, University of Tennessee, Knoxville, TN, USA}
\author{Jose L. Lado}
\email{jose.lado@aalto.fi}
\affiliation{Department of Applied Physics, Aalto University, Espoo, Finland}

\date{\today}

\begin{abstract}
Tunneling Andreev reflection (TAR) spectroscopy provides a new approach to identify superconducting pairing symmetry at the atomic scale. Using atomistic superconducting transport simulations, we reveal the mechanism by which TAR can distinguish between pairing symmetries, which is complementary to traditional conductance-based techniques. In particular, owing to the additivity of the excess tunneling decay rate, the TAR spectrum is a weighted average of the contributions from quasiparticle currents, Andreev reflection, and higher-order scattering processes, and their relative weights depend on both the superconducting order parameter and the coupling strength. Within the local tunneling model, TAR dominates mid-gap conductance for s-wave superconductors, is suppressed for d-wave, where the momentum-averaged superconducting gap vanishes, and coexists with quasiparticle tunneling in sign-changing symmetries for which it remains finite. Meanwhile, higher-order processes generally enhance the TAR signal when $G_N$ exceeds approximately $0.1G_0$. As a result, TAR provides a rich spectral fingerprint of the underlying pairing symmetry and electronic structure, enabling atomically resolved identification of unconventional superconducting states.
\end{abstract}

\maketitle
\section{Introduction}
Pairing symmetry provides crucial insights into superconductors \cite{Tsuei2000, Chubukov2012,RevModPhys.63.239,Yin2014} and can reveal the presence of exotic phenomena like topological superconductivity\cite{Flensberg2021,RevModPhys.83.1057,Sato2017,Alicea2012}. Many techniques such as muon spin relaxation ($\mu \mathrm{SR}$) \cite{Hillier2022, Blundell2025}, quasiparticle interference imaging \cite{Ming2023,Hnke2012,Gao2018,Wang2025} and point contact Andreev reflection (PCAR) \cite{Diez2012, Deutscher2005, Asano2021} have contributed significantly to understanding the symmetry of the order parameter in conventional and unconventional superconductors. However, the definitive assignment remains challenging and often controversial for any new material, stimulating the search for new, complementary, and more direct methods\cite{2025arXiv250922902W,Khosravian2024,PhysRevB.111.014501}. Among them, techniques that exhibit high spatial resolution are particularly useful for addressing the properties of inhomogeneous superconductors and superconducting interfaces enabling the creation of emerging electronic states \cite{Shibauchi2020,PhysRevB.108.024505,Liebhaber2022,Schneider2022,Jiao2020,Kezilebieke2020,Huang2017,Liu2021}.

\begin{figure*}[t]
    \centering
    \includegraphics[width=\linewidth]{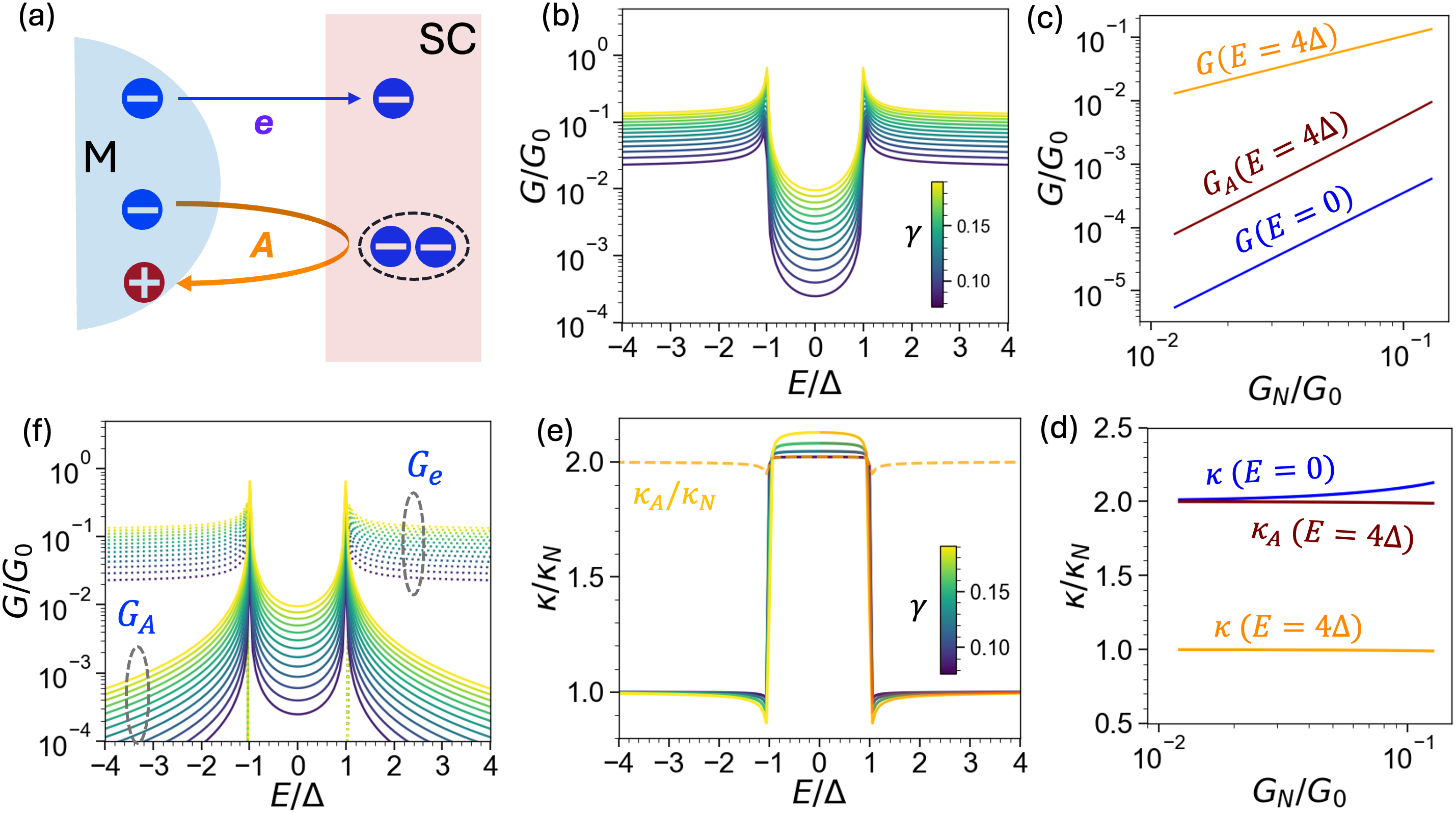}
    \caption{Simulated conductance and $\kappa$-spectra for s-wave superconductor at weak to intermediate coupling strength.  (a) Schematic of single quasiparticle tunneling (top) and Andreev reflection (bottom) between a metallic tip and a superconducting sample. (b) Conductance spectroscopy for a range of coupling.  (c) Conductance scaling  as a function of the normal conductance, $G_N/G_0$ (measured outside the superconducting gap), for the total conductance outside ($E=4\Delta$) and inside ($E=0$) the gap, and for the Andreev component outside the gap ($G_A$). (d) $\kappa/\kappa_{N}$ calculated for energies inside the gap $(E=0)$, outside the gap ($E=4\Delta$) and separately for Andreev reflection outside the gap ($\kappa_A$). (e) Excess decay rate ($\kappa/\kappa_{N}$) spectra calculated from (b). Red highlights spectral reconstruction using Eq.~\ref{eq:kappaadd} from the main text. Dashed line shows the scaling of only the Andreev reflection channel. (f) Partitioning of the conductance data in (b) into contributions from tunneling (dotted) and Andreev reflection (solid).   Colorscales in (b), (d) and (e) follow the values of the coupling strength.  $G_0$ is the conductance quantum.}
    \label{fig:fig1}
\end{figure*}

\begin{figure*}[t]
    \centering
    \includegraphics[width=\linewidth]{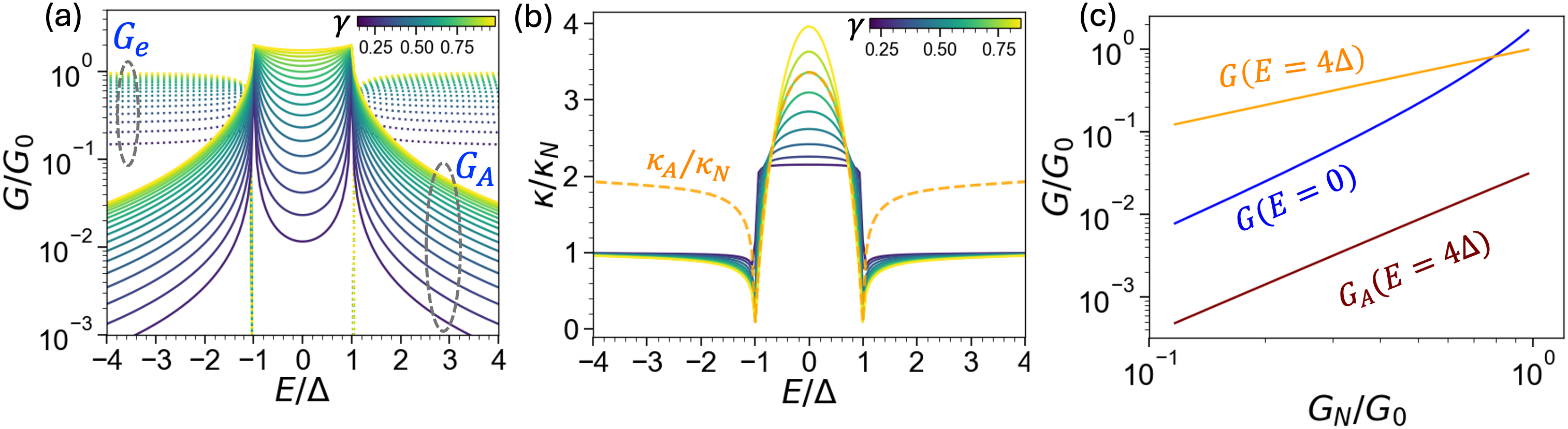}
    \caption{Near-contact conductance and $\kappa$-spectra for s-wave superconductor at strong coupling: (a) Conductance spectroscopy as a function of increasing coupling strength partitioned into contributions from tunneling (dotted) and Andreev reflection (solid). (b) Corresponding $\kappa/\kappa_{N}$ spectroscopy; the orange dashed line shows $\kappa_A/\kappa_{N}$, the scaling of the Andreev reflection channel alone. Most notably, $\kappa/\kappa_N$ significantly exceeds the anticipated value of 2, and can reach as high as 4 as the mid-gap conductance approaches 2$G_0$. (c) Scaling of the total conductance outside ($E=4\Delta$) and inside ($E=0$) the gap, and of the Andreev component outside the gap ($G_A$), as a function of normal conductance (measured outside superconducting gap). Colorscales in (a) and (b) reflect
    the different tip-substrate coupling strength $\gamma$.   $G_0$ is the conductance quantum.}
    \label{fig:fig2}
\end{figure*}

\begin{figure*}[t]
    \centering
    \includegraphics[width=\linewidth]{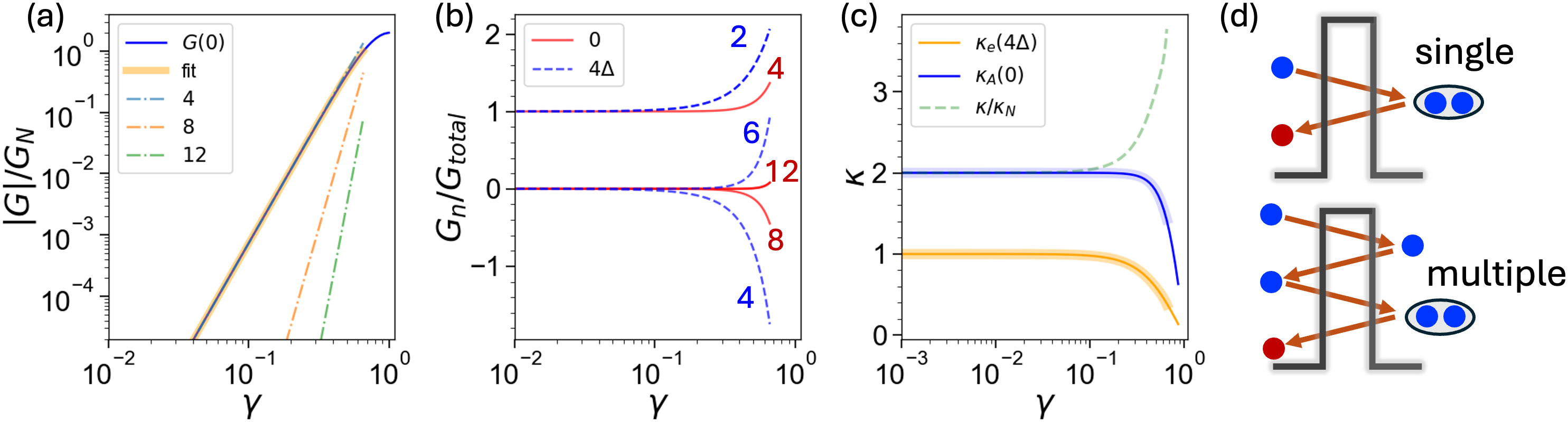}
\caption{  (a) Polynomial fitting of the calculated conductance
	within the BTK formalism. Dashed-dotted lines show absolute value of
	conductance from individual polynomial terms, with order sequence from
	Eq.~\ref{eq:taylor1}
	for mid-gap conductance and Eq.~\ref{eq:S4}
	for above-gap conductance.
	The sum of these contributions is overlaid using orange shading
	providing an excellent fit to net conductance. (b) Ratio of conductance
	of the individual polynomial terms ($G_n$) of order n obtained from (a)
	and Fig.~\ref{fig:S1}, including their sign. Blue dashed and red lines correspond to
	above and mid-gap conductance, respectively. The numbers mark the
	order of the specific polynomial. (c) 	$\kappa$ as a function of
	$\gamma$ for above-gap tunneling (orange) and mid-gap Andreev
	reflection (blue curve). Both gradually reduce with increased coupling,
	particularly when $\gamma > 0.1$. However, the detailed dependence is
	distinct between them. The shaded lines are fits obtained from the
	conductance data using a weighted
	average Eq.~\ref{eq:kappaadd}, with
	weighting from (b). The ratio of the $\kappa$ values is shown as green
	dotted line.  (d) Schematic of semiclassical trajectories corresponding
	to single and multiple (higher order) Andreev reflection.
}

    \label{fig:fig3}
\end{figure*}

\begin{figure*}[t]
    \centering
    \includegraphics[width=\linewidth]{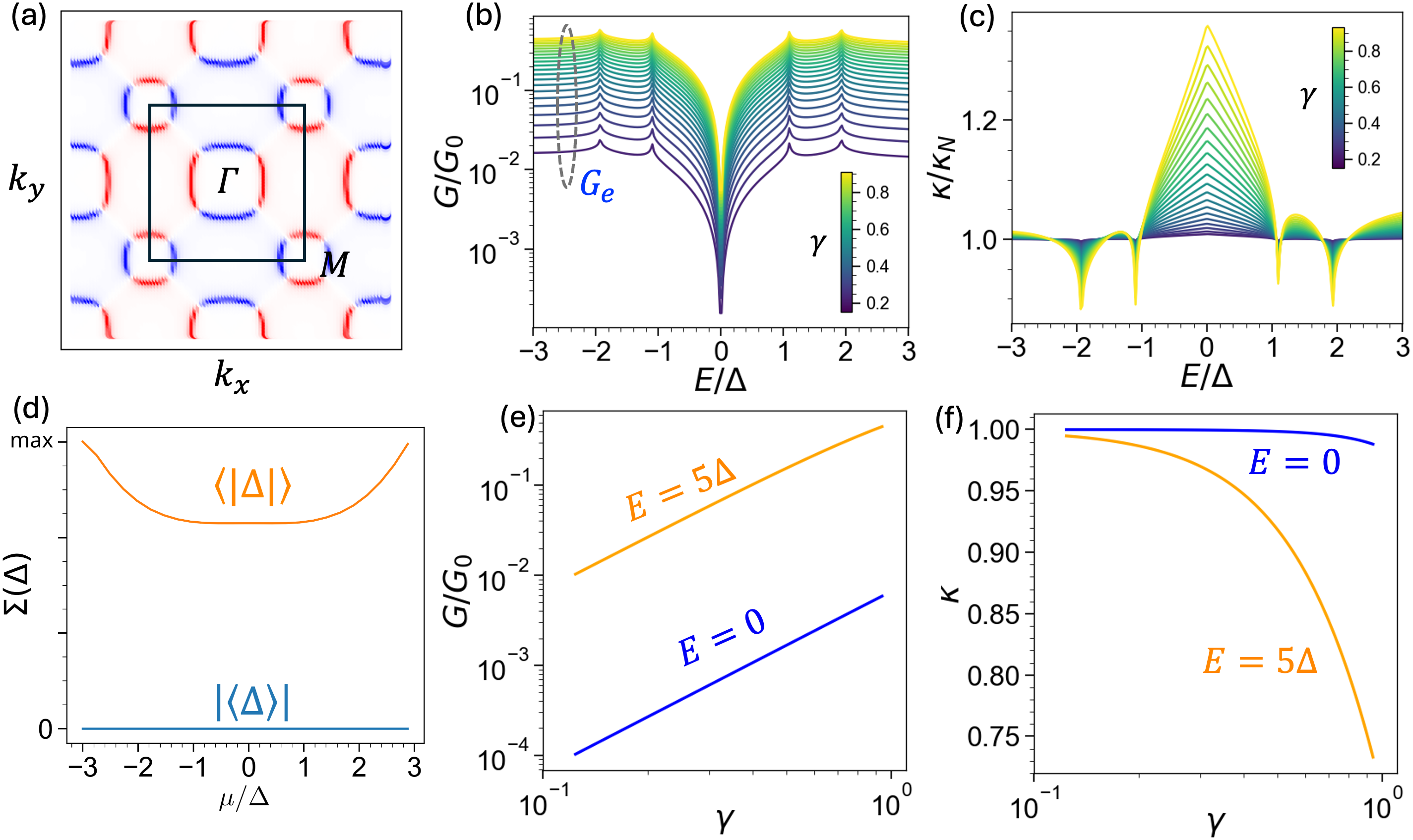}
    \caption{$\kappa$-spectroscopy for a d-wave superconductor, where  tunneling Andreev reflection is suppressed: (a) Fermi surface for a two-Fermi-surface electron structure with the d-wave order parameter. (b) Conductance spectroscopy as a function of increasing coupling strength. The Andreev contribution to this conductance is below numerical accuracy, so that calculated conductance is from single quasiparticle tunneling at any energy. (c)  $\kappa/\kappa_{N}$ spectra calculated from conductance in (b). (d) The expectation value of the superconducting gap function ($|\langle \Delta \rangle|$, as defined in the main text) is zero for any chemical potential (blue curve), while the average absolute value of the SC gap ($\langle | \Delta | \rangle$) remains finite. The axes are the spectral-weighted gap average $\Sigma(\Delta)$ defined in the main text, versus the chemical potential $\mu$ at which it is evaluated. (e) Conductance and (f) unnormalized decay rate $\kappa$
    as a function of coupling strength $\gamma$,
    at energies in the middle ($E=0$) and outside ($E=5\Delta$) the superconducting gap.  Mid-gap (blue) conductance saturates slightly slower than outside the gap (orange) when approaching contact, which is effectively captured by the slight differences of the unnormalized $\kappa$ (f) and $\kappa/\kappa_{N}$ spectroscopy (c). Colorscales in (b) and (c) follow the values of the coupling strength $\gamma$.}    
    \label{fig:fig4}
\end{figure*}

\begin{figure*}[t]
    \centering
    \includegraphics[width=\linewidth]{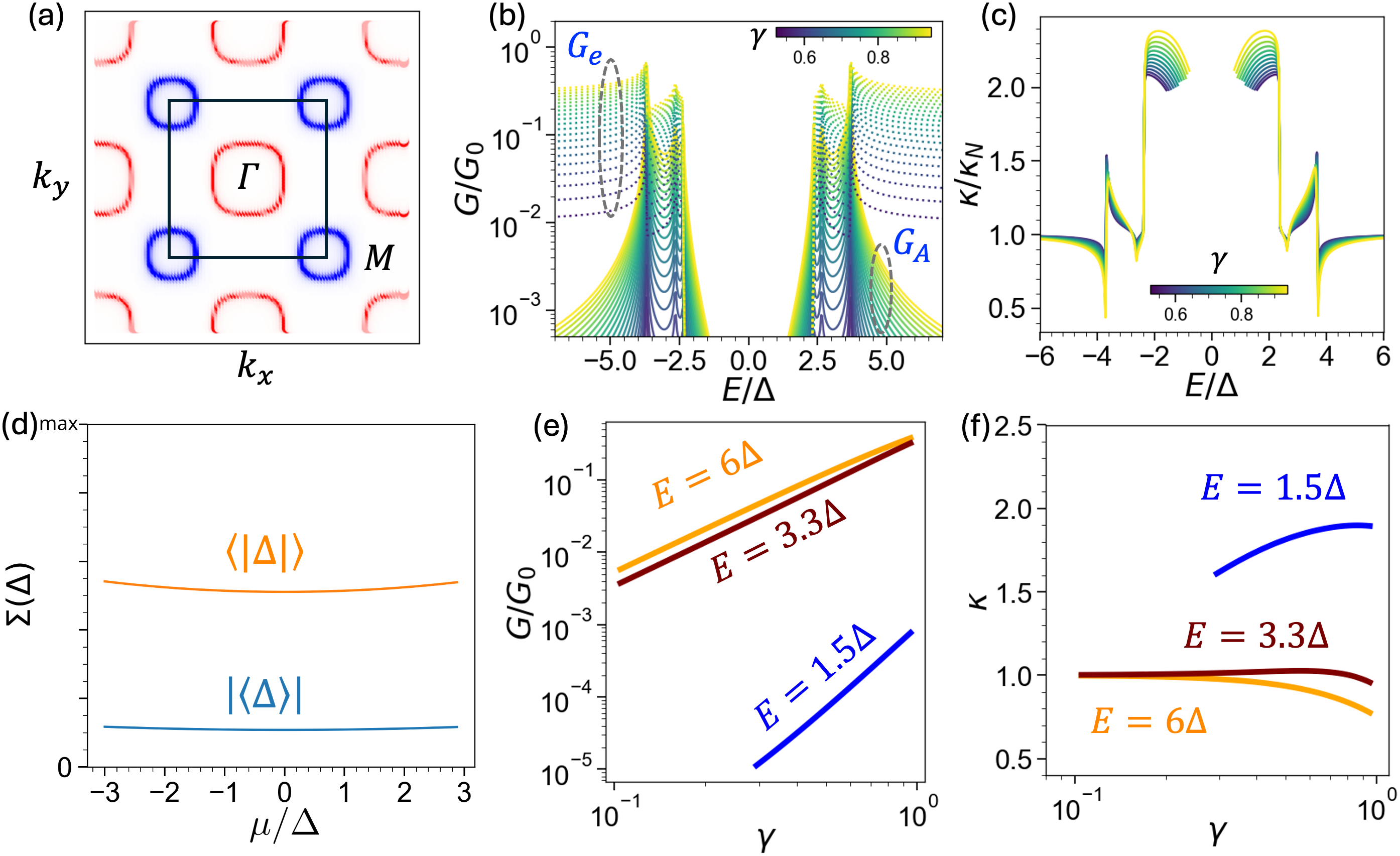}
    \caption{$\kappa$-spectroscopy for $s_{\pm}$ superconductor, where there exists competition between quasiparticle tunneling and Andreev reflection as a function of energy: (a) Fermi surface of a two-Fermi-surface superconductor with the $s_{\pm}$ superconducting order parameter. (b) Conductance spectroscopy as a function of increasing coupling strength. Solid lines are the Andreev component ($G_A$), and dotted lines are quasiparticle tunneling ($G_e$); (c)  $\kappa/\kappa_{N}$ spectra calculated from conductance in (b). (d) Expectation value of both superconducting gap ($|\langle \Delta \rangle|$) and gap magnitude $\langle |\Delta |\rangle$ remain finite for any chemical potential; the axes are the spectral-weighted gap average $\Sigma(\Delta)$ defined in the main text, versus the chemical potential $\mu$ at which it is evaluated. Scaling of conductance (e) and unnormalized decay rate $\kappa$ (f)
    as a function of the coupling strength $\gamma$, at several energies across the gap.} 
    \label{fig:fig5}
\end{figure*}
\begin{figure*}[t]
\centering
\includegraphics[width=0.72\linewidth]{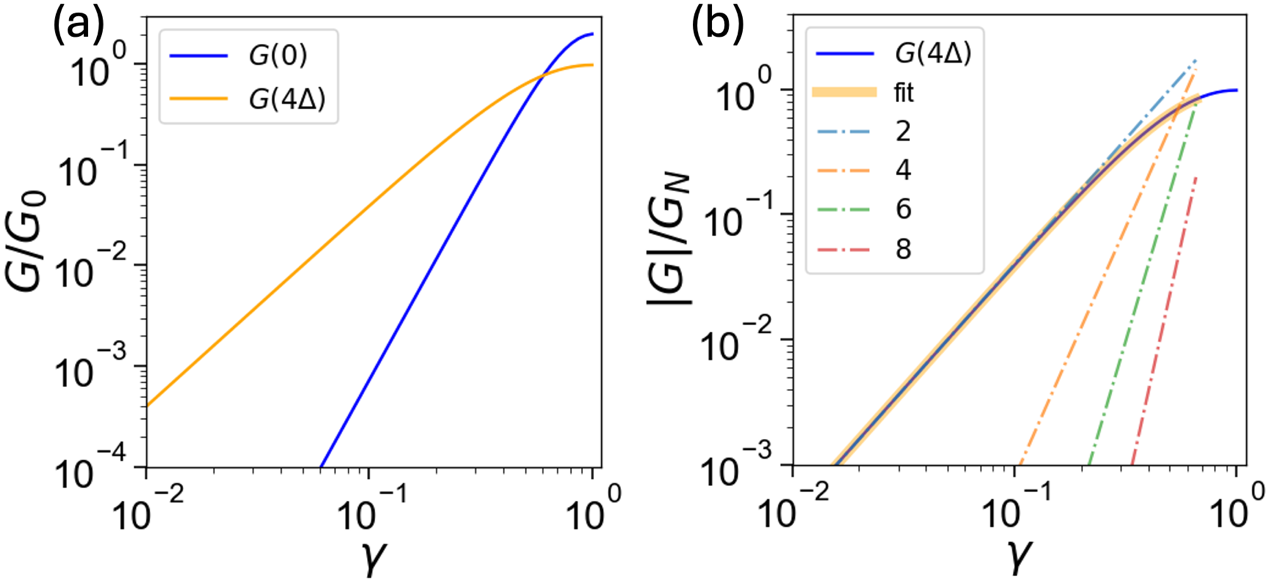}
\renewcommand{\thefigure}{S1}
\caption{(a) Mid-gap (blue) and above-gap (orange) conductance vs.\ coupling strength ($\gamma$) numerically calculated for the s-wave superconductor. (b) Polynomial fitting of the calculated above-gap conductance within the BTK formalism. Dashed-dotted lines show absolute value of conductance from individual polynomial terms, with order sequence from Eq.~\ref{eq:S4}. The sum of these contributions is overlaid using orange shading.}
\label{fig:S1}
\end{figure*}
\begin{figure*}[t]
\centering
\includegraphics[width=\linewidth]{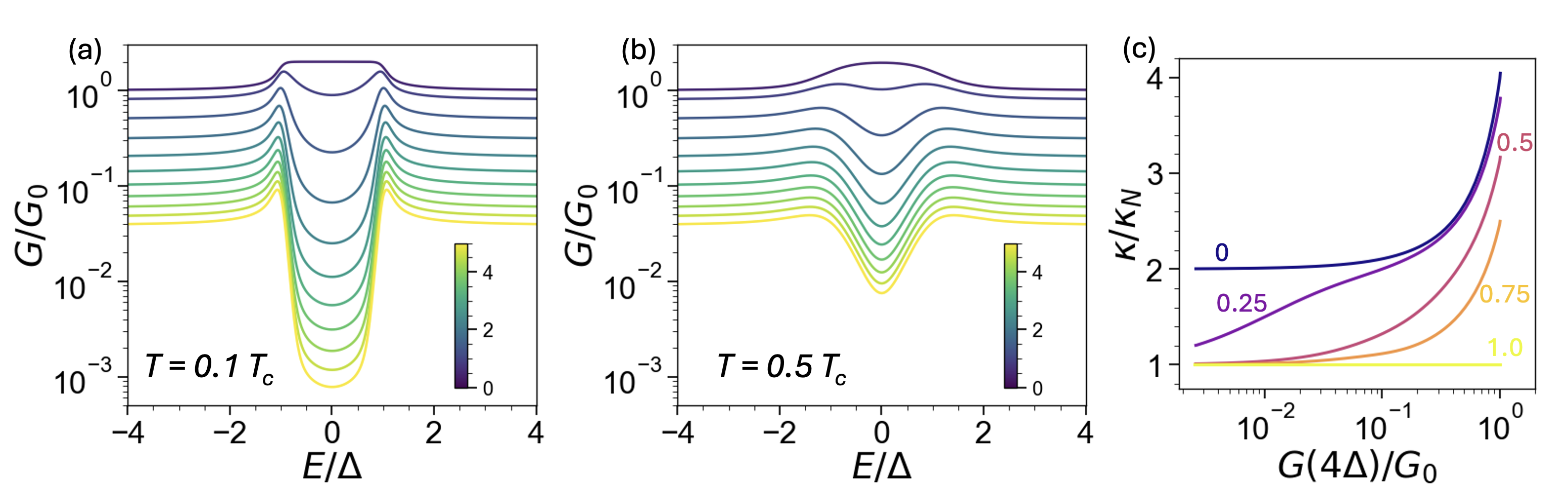}
\renewcommand{\thefigure}{S2}
\caption{(a,b) Conductance spectra for s-wave superconductor at $T = 0.1\,T_{c}$ and $T = 0.5\,T_{c}$ calculated using BTK theory incorporating finite temperature. (c) $\kappa/\kappa_{N}$ scaling with coupling strength for increasing temperature (from $T = 0$ to $T = 1.0\,T_{c}$) marked on the plots.}
\label{fig:S2}
\end{figure*}

Tunneling Andreev Reflection (TAR) provides a new approach to detect Andreev reflection in scanning tunneling microscopy (STM) with near atomic-scale resolution \cite{Ko2022,Ko2023}. Whereas traditional point contact Andreev spectroscopy detects excess conductance as a proxy for Andreev reflection across a metal-superconductor junction, TAR relies on the excess tunneling decay rate ($\kappa$). Methodologically, STM probes tunneling currents at different tunneling couplings ($\gamma$) to precisely measure the coupling dependence of the tunneling current ($G \sim |\gamma|^{2\kappa}$). Andreev and electron components of the tunneling current can then be disentangled by comparing the exponent $\kappa$ of the power-law dependence in the superconducting state ($\kappa$) vs normal state ($\kappa_{N}$). The strong dependency of Andreev reflection current on the tunneling coupling is well known and is generally considered to be detrimental due to rapid suppression of the excess conductance \cite{Deutscher2005, Asano2021, Meyer2025, Brand2017, Sukhachov2023}. By contrast, this very same dependency is directly measured in TAR so that this technique can, in principle, operate at arbitrarily small contact transparency. 

Understanding the factors that contribute to specific shapes of the $\kappa$-spectra in TAR remains an open problem. For the simplest case of an s-wave superconductor, the ratio $\kappa / \kappa_{N}$ equals 2 inside the gap. This ratio directly reflects a simple intuitive picture of two particles, an electron and a retroreflected hole, traversing the junction in Andreev reflection \cite{Cuevas2003, Johansson2003}, with net probability of Andreev reflection being a product of individual tunneling events. However, $\kappa / \kappa_{N}$ can readily deviate from 2 \cite{Ko2022}. Even for the s-wave, $\kappa / \kappa_{N}$ can actually increase beyond $2$ at smaller barriers. Meanwhile, for more complicated $\mathrm{d}, \mathrm{p}, \text{and } \mathrm{s}_{\pm}$ symmetries, the $\kappa$-spectra 
generally exhibit a complex energy dependence \cite{Ko2023}. These rich details enable ``fingerprinting'' superconductors, and ultimately connecting the observations with specific symmetries. It is therefore imperative to establish the fundamental origins of these spectra and further identify signatures of TAR in unconventional superconductors.

Here, we develop a quantitative interpretation of the $\kappa$-spectroscopy based on direct calculation of Andreev reflection and electron tunneling. The structure of the $\kappa$-spectra is determined by energy-dependence of Andreev tunneling, the expectation value of the superconducting gap across the Brillouin zone, the competition between Andreev and single quasiparticle tunneling and higher order tunneling and Andreev reflection processes at stronger coupling. The varying contribution of these factors provides a clear path to explain and differentiate several representative pairing symmetries. We also provide a quantitative analysis of the near-contact regime where higher order tunneling terms provide the strongest signatures of Andreev reflection and can even provide a unique
$\kappa$-spectrum for a superconductor where tunneling Andreev reflection is suppressed by symmetry. 

\section{Methods}

The calculated conductance can be decomposed into contributions of Andreev reflection and single quasiparticle tunneling, 
which is particularly important if several mechanisms can simultaneously contribute to conductance.
While such a decomposition cannot be performed experimentally,
the tunneling-dependence of the current will enable separating both contributions.
We model the STM substrate system using a generalized tight binding model of the form

\begin{equation}
\mathcal{H} = \mathcal{H}_{\text{tip}} + \mathcal{H}_{\text{substrate}} + \mathcal{H}_{\text{tunneling}} 
\end{equation}
where we take a normal metal tip with Hamiltonian
\begin{equation}
\mathcal{H}_{\text{tip}} = t \sum_{n>0,s} \left ( d^\dagger_{n,s} d_{n+1,s} + h.c. \right )
\end{equation}
where $d^\dagger_{n,s}$ creates an electron on tip site $n$ with spin $s$.
The two-dimensional superconducting substrate Hamiltonian takes the form
\begin{equation}
\mathcal{H}_{\text{substrate}} =  \sum_{ijs} t_{ij} c^\dagger_{i,s} c_{j,s} + \sum_{ijss'} \left ( \Delta^{ss'}_{ij} c^\dagger_{i,s} c^\dagger_{j,s'} + h.c. \right )
\end{equation}
where $c^\dagger_{i,s}$ creates an electron on the substrate site $i$ with spin $s$.
We consider a square lattice system with long range hoppings $t_{ij}$,
whose specific strengths allow us to create single and multiple Fermi surfaces.
The superconducting order $\Delta^{ss'}_{ij}$ allows us to include both
conventional and unconventional superconducting orders within the same model, and
specifically s-wave,
extended s-wave and d-wave.
The tunable Hamiltonian coupling the tip
and substrate takes the form
\begin{equation}
\mathcal{H}_{\text{tunneling}} = \gamma
\sum_s c^\dagger_{0,s}d_{0,s} + h.c.
\end{equation}
The tunneling junction is accounted for by the variable hopping $\gamma$
connecting metallic tip and superconducting substrate
(see schematic in Fig.~\ref{fig:fig1}a). 
From an experimental perspective, the coupling $\gamma$ shows
an exponential dependence as a function of the distance between tip and sample. 
The tunneling barrier in these calculations is highly localized in space, thereby precluding momentum resolution of tunneling current. This condition approximates STM experiments with atomic-scale contacts \cite{Hofer2003}. In the tunneling regime,
the conductance is proportional to $|\gamma|^2$.

To disentangle the Andreev and quasiparticle contribution,
the conductance will be
separated into Andreev reflection and single quasiparticle tunneling via decomposition of the scattering matrix into
\begin{equation}
G (\omega,\gamma) = G_{e} (\omega,\gamma) + G_{A} (\omega,\gamma)
\end{equation}
where $G_{e}$ corresponds to single electron transmission and $G_{A}$
corresponds to double electron transmission, known as Andreev reflection, where an incoming electron gets reflected as a hole.
The previous conductances are computed using an $\mathcal{S}$-matrix formalism implemented
with Nambu non-equilibrium Green's functions \cite{PhysRevB.23.6851,PhysRevB.97.195429,Sancho1985,datta1995electronic,pyqula}, where the tip is taken as quasi one dimensional and the substrate as an infinite two-dimensional plane.
It is worth noting that, although this geometry breaks the translational symmetry of the
substrate due to the STM tip, the exact $\mathcal{S}$-matrix can be computed using
embedded Green's functions \cite{PhysRevB.96.024403,PhysRevMaterials.6.094010}.
The tunneling dependence of the conductance allows us to define the decay rate as
\begin{equation}
\kappa (\omega,\gamma) = \frac{1}{2} \frac{d \log G (\omega,\gamma)}{d \log \gamma} = \frac{1}{2} \frac{\gamma}{G(\omega,\gamma)}\frac{dG (\omega,\gamma)}{d\gamma}
\label{eq:kappa}
\end{equation}
The interpretation of the previous object can be rationalized from the power series of the conductance. The conductance $G$ will have a dominating power dependence as a function
of $\gamma$ of the form $ G \sim \alpha |\gamma|^\beta$ for $\gamma \ll 1$. For single electron tunneling, we expect $\beta=2$ and $\kappa=1$, whereas for Cooper pair tunneling in a conventional superconductor $\beta=4$ and $\kappa=2$, respectively. As a result, the decay rate enables one to identify from the tunneling-dependent conductance the presence or absence of Cooper pair tunneling. From Eq.~\ref{eq:kappa} 
it can be readily observed that the decay rate extracts specifically
the power dependence $\beta$ in the small coupling regime $\gamma \ll 1$. 
In contrast, for stronger coupling, 
$\kappa$ signals the presence of higher order power
expansions of the conductance. Finally, it is convenient to define the electron and
Andreev decay rates as
\begin{equation}
\kappa_e (\omega,\gamma) = \frac{1}{2} \frac{\gamma}{G_e(\omega,\gamma)}\frac{dG_e(\omega,\gamma)}{d\gamma}
\end{equation}
\begin{equation}
\kappa_A (\omega,\gamma) = \frac{1}{2} \frac{\gamma}{G_A(\omega,\gamma)}\frac{dG_A(\omega,\gamma)}{d\gamma}
\end{equation}

From the definition of Eq.~\ref{eq:kappa}, it is straightforward to show that $\kappa$ is a weighted average of the two contributions:

\begin{equation}
\kappa (\omega,\gamma) =\kappa_{e} (\omega,\gamma) \frac{ G_{e}(\omega,\gamma)}{G(\omega,\gamma)}+\kappa_A(\omega,\gamma)\frac{G_{A}(\omega,\gamma)}{G(\omega,\gamma)}
\label{eq:kappaadd}
\end{equation}
where the total conductance $G=G_{e}+G_{A}$ equals the sum of quasiparticle and Andreev conductances (i.e. a parallel circuit). The additive property is a direct consequence of the definition of $\kappa$ and the additivity of conductance of independent contributing channels.
For convenience, in the following we will normalize the decay rates to the decay rate at energies
well above the superconducting gap $\kappa_N$, which would correspond to single electron transmission. We note that experimentally $\kappa$ is calculated in a similar approach, but using above-gap conductance as a proxy for the coupling strength \cite{Ko2023}. To minimize noise, $\kappa$ is measured by fitting the $\log G(E) = f( \log G_N)$ dependency at a given energy with a first or second order polynomial.

\section{Results and Discussion}
We first revisit the simplest case of tunneling into an s-wave superconductor. Fig.~\ref{fig:fig1}b presents the variation of the calculated conductance spectrum at $\mathrm{T}=0 \mathrm{~K}$ as a function of the coupling strength. The largest normal conductance (at $E=4 \Delta$) in Fig.~\ref{fig:fig1}b is $0.13 \mathrm{G}_{0}$, so all the spectra remain essentially in the tunneling regime; the crossover to the near-contact regime around $G_{N} \sim 0.1 G_{0}$ is gradual rather than sharp. Fig.~\ref{fig:fig1}c further compares mid-gap (blue) and above-gap (orange) conductance. It is quite evident from these figures that the conductance grows much faster inside the superconducting gap as a function of increased coupling. The excess decay rate ($\kappa / \kappa_{N}$) spectra calculated from the conductance spectra are shown in Fig.~\ref{fig:fig1}d and e. As we discussed previously \cite{Ko2022}, $\kappa / \kappa_{N} \geq 2$ in the middle of the gap. This spectrum can be calculated at arbitrarily small coupling strength (limited by computational accuracy or noise in the case of the experiment), which is one of the key strengths of the TAR approach. Note that the calculation of $\kappa (\gamma)$ depends
on the coupling between tip and sample $\gamma$, and thus can be fundamentally different in the tunneling $G_N \ll G_0$ and near contact $G_N \sim G_0$ regimes.

Partitioning the total conductance into contributions due to Andreev reflection ($G_{A}$) and quasiparticle tunneling ($G_{e}$), as shown in Fig.~\ref{fig:fig1}f, provides a crucial step to understanding the structure of any $\kappa$-spectra. For a fully gapped superconductor at 0 K, the quasiparticle current inside the superconducting gap is zero and the conductance is entirely due to Andreev reflection \cite{Deutscher2005, Asano2021}. Meanwhile, outside the superconducting gap the Andreev current (solid lines) coexists with quasiparticle tunneling (dotted lines) (Fig.~\ref{fig:fig1}f), and quickly subsides with increasing energy. 

A key property of $\kappa / \kappa_{N}$ is its additivity as shown in Eq. \ref{eq:kappaadd}. Figure~\ref{fig:fig1}e overlays $\kappa / \kappa_{N}$ calculated using Eq.~\ref{eq:kappaadd} and the Andreev conductance ratio. 
Clearly, the $\kappa / \kappa_{N}$ calculated from separate channels is identical to one calculated from total conductance, which verifies the additivity of the $\kappa$. The overall picture at small $G_{N}$ is now clear: at 0 K the conductance mechanism is exclusively Andreev reflection inside the superconducting gap. Above the gap energy, Andreev reflection and quasiparticle tunneling coexist. However, the conductance of the Andreev channel is very rapidly suppressed with increasing energy as well as decreasing coupling (compare brown and orange lines in Figure~\ref{fig:fig1}c), so that the conductance outside the superconducting gap is almost entirely due to quasiparticle tunneling. As a result, the $\kappa / \kappa_{N}$ spectrum is essentially a 1 to 2 transition, with few other significant features in the spectra. One interesting caveat shown in Fig.~\ref{fig:fig1}d is that outside the superconducting gap, the excess decay rate due to Andreev tunneling is almost exactly two (brown line), while it is slightly larger than two inside the gap (blue line). We will analyze this property in more detail below.

We now turn to the near-contact regime, where $G_{N}$ approaches $G_{0}$. Conductance spectra in this regime begin to rapidly fill the spectral
gap (Fig.~\ref{fig:fig2}a) due to Andreev reflection (solid lines), and then eventually the mid-gap conductance reaches approximately twice
the values outside the gap, a classic signature of Andreev reflection
\cite{Deutscher2005, Asano2021}. The $\kappa$-spectrum likewise deviates from
the weak coupling in this regime. Specifically, $\kappa / \kappa_{N}$ can reach
values as high as 4 mid-gap (Fig.~\ref{fig:fig2}b) in the window
$G_{0}>G_{N}>0.1 G_{0}$. This has been referred to as near-contact Andreev
reflection \cite{Ko2022, Ko2023}. The growth rate of Andreev conductance with
increased coupling (blue in Fig.~\ref{fig:fig2}c and Fig.~\ref{fig:S1}a ) clearly
becomes larger in the near-contact region. Eventually Andreev conductance
mid-gap exceeds that of junction conductance outside the gap (orange in
Fig.~\ref{fig:fig2}c and Fig.~\ref{fig:S1}a), ultimately leading to a 2:1 conductance ratio at contact. Notably, Andreev conductance outside the gap (brown line in
Fig.~\ref{fig:fig2}c) does not exhibit this effect. 

Using the additivity of $\kappa / \kappa_{N}$, we can explain the near-contact
Andreev reflection by allowing multiple reflections in the junction, where for
each Andreev reflection, quasiparticles scatter more than twice in the junction
region, conceptually similar to Refs. \cite{Asano2021,Beenakker2000}.
The starting point is a closed form expression for the sub-gap conductance from the BTK theory \cite{Blonder1982}, expressed as a series in powers of coupling strength $\gamma$:

\begin{equation}
G_A(E = 0) \approx 8\gamma^4 - 16\gamma^8 + 24\gamma^{12} - ...
	\label{eq:taylor1}
\end{equation}

wherein Eq.~\ref{eq:taylor1} is derived from Eq.~\ref{eq:S3} by setting $E = 0$ (see Appendix~\ref{app:btk} for more details). Eq.~\ref{eq:taylor1} and a similar expression for above-gap conductance (Eq.~\ref{eq:S4}) can be interpreted as a net conductance of parallel conductors, each represented by a
polynomial term. The leading order term of Andreev reflection (Eq.~\ref{eq:taylor1}) scales as
square of tunneling (above gap) conductance (Eq.~\ref{eq:S4}), as expected. We can now use the
sum of respective polynomial terms to fit the conductance as a function of
coupling strength (Fig.~\ref{fig:fig3}a). The fits clearly show the prevalence
of lowest order term for Andreev
reflection (Fig.~\ref{fig:fig3}a) as well as tunneling (Fig.~\ref{fig:S1}b) at weak coupling ($\gamma < 0.1$) and the
increased contribution of higher order terms in the near-contact window
($\gamma > 0.1$). This is best seen by plotting the conductance ratio for each of the fitted contributions as a function of coupling (Fig.~\ref{fig:fig3}b). Note that the sign of the terms in Eq.~\ref{eq:taylor1} and Eq.~\ref{eq:S4} alternates,
so that their relative contribution can either enhance or suppress the
total conductance. Finally, we can use Eq.~\ref{eq:kappaadd} to calculate the decay rate
$\kappa$ as a sum of contributions from each of the fitted polynomial terms,
weighted by their conductance ratio (Fig.~\ref{fig:fig3}b). The fits are 
excellent for both above-gap tunneling and mid-gap Andreev reflection
(Fig.~\ref{fig:fig3}c), reproducing both behaviors at weak and stronger
coupling. The ratio of the two, $\kappa/\kappa_N$, exceeds $2$ when $\gamma > 0.1$ and can exceed $3$ in this window. 

The emerging picture of the near-contact window is that both $\kappa_e$ and $\kappa_A$ decrease
with increased coupling  (Fig.~\ref{fig:fig3}c), reflecting the effective reduction of the tunneling barrier under the image potential
\cite{Olesen1996}. The higher order processes, which generally involve multiple
barrier crossings (Fig.~\ref{fig:fig3}d), become more effective in this regime.
The specific structure of higher order contributions is different for
Andreev reflection and tunneling. As a result, $\kappa/\kappa_{N}$ deviates
from its value of $2$ characteristic of the lowest order processes, and this
deviation marks the involvement of higher order processes in the
measured conductance. 

The higher order scattering should not be confused with multiple Andreev
reflection (MAR) in a superconducting junction \cite{Kleinsasser1994, Cuevas2003,
Johansson2003}. Instead, it is a single Andreev reflection that proceeds
through a sequence of multiple reflections, schematically shown in
Fig.~\ref{fig:fig3}(d). The existence of such processes has been discussed
several times \cite{Beenakker2000, Asano2021}. However, to our knowledge
$\kappa$-spectroscopy provides the first direct method to probe them, owing to
the additivity of the $\kappa$ as an observable.

From the application point of view, the near-contact window provides the
strongest signals for tunneling Andreev reflection. For example, at finite
temperature, the tunneling of thermally excited quasiparticles competes with
Andreev reflection and will reduce $\kappa/\kappa_{N} < 2$ at all energies (see Appendix~\ref{app:thermal}). Due to additional effect of higher order terms, the near-contact signal persists to much higher
temperatures compared to weak coupling, ultimately enabling detecting Andreev reflection and a fingerprint of superconductivity almost to the transition temperature. Of course, the natural limitation for these measurements is the stability of a proximate junction, which will depend on specific materials and measurement conditions.

The analysis for the s-wave order parameter provides the fundamental ingredients needed to interpret the $\kappa$-spectra of nodal and sign-changing superconductors. In the following we will discuss $d_{x^{2}-y^{2}}$ and $s_{ \pm}$ pairing symmetry on a 
Hamiltonian with two Fermi surfaces, previously used to describe
FeSe \cite{Ko2023}. Figure~\ref{fig:fig4} shows the analysis of $\kappa$-spectra for the $d_{x^{2}-y^{2}}$ order parameter with two-Fermi-surface Hamiltonian (Fig.~\ref{fig:fig4}a). The conductance and the corresponding $\kappa / \kappa_{N}$ spectra are shown in Fig.~\ref{fig:fig4}b and c, respectively. As we have previously noted, the difference between s- and d-wave is very large. $\kappa / \kappa_{N}$ remains essentially equal to unity across the superconducting gap (Fig.~\ref{fig:fig4}c). Only in the near-contact regime with $G_{N}>0.1 G_{0}$ the spectra begin to show structure. $\kappa / \kappa_{N}$ values grow up to $\approx 1.35$ in the middle of the gap, and have a pronounced central peak structure that can be qualitatively described as the ``inverse'' of the gap in the conductance spectra (Fig.~\ref{fig:fig4}c).

It turns out that none of these features correspond to Andreev reflection. The
calculated Andreev current is negligible compared to tunneling current. A complete
suppression of Andreev reflection is rooted in the sign change of the order
parameter across the Brillouin zone and an equal volume of the positive and
negative gap values. The detailed calculation of Andreev reflection
amplitude needs to consider the specific structure of the coupling elements
between the probe and the superconducting surface.
In practice, these are given by the hybridization
functions\cite{PhysRevB.23.6851} $\int \gamma_{\mathbf k} 
(\omega-H_{\mathbf k}+i0^+)^{-1} \gamma^\dagger_{\mathbf k} d^2 \mathbf k 
$ ,
where the off-diagonal element in the Nambu basis encodes
the amplitude of the Andreev reflection,
$H_{\mathbf k}$ is the Nambu Hamiltonian
of the substrate,
and $\gamma_{\mathbf k}$ is the momentum-dependent tip-surface
coupling.
Under an assumption of momentum-independent
coupling strength $\gamma_{\mathbf k}$, 
the effectiveness of Andreev reflection stems
from the amplitude of the off-diagonal
element of the Nambu Green's function,
which to first order
is proportional to the spectral-weighted momentum
average of the gap,
\begin{equation*}
\langle \Delta \rangle (\mu) \equiv
\frac{\int \Delta(\mathbf k) A(\mathbf k,\omega=0,\mu) \, d^2 \mathbf k}
     {\int A(\mathbf k,\omega=0,\mu) \, d^2 \mathbf k}
\end{equation*}
where $A(\mathbf k,\omega=0,\mu)$ is the momentum-resolved spectral function of
the substrate evaluated at the Fermi energy $\omega=0$, for a finite chemical
potential $\mu$. For $d_{x^{2}-y^{2}}$ this average vanishes for any chemical
potential (Fig.~\ref{fig:fig4}d). 

The structure of the $\kappa$-spectra is then related to the
impact of superconductivity in the electronic spectral function,
and arises from quasiparticle tunneling. Here too, it is helpful to compare the coupling-dependence of conductance (Fig.~\ref{fig:fig4}e) and $\kappa$ (Fig.~\ref{fig:fig4}f) mid-gap (blue) and well outside the gap (orange). The decay rate in the middle of the gap reduces slightly slower with increased coupling. And the $\kappa$-spectra sensitively capture this relative difference as a function of energy (Fig.~\ref{fig:fig4}c). We propose that this effect is also a signature of higher-order tunneling processes, and it is rooted in the decrease of the density of states inside the superconducting gap. We have verified that a similar result is observed for a single band Hamiltonian with d-wave symmetry. And in principle this effect should manifest for any electronic gap structure. From the perspective of identifying pairing symmetry, suppression of Andreev reflection in the tunneling junction provides a ``litmus test'' for finite angular momenta Cooper pairing, and 
in particular d-wave superconductivity, compared to s-wave order parameter.

The connection between tunneling Andreev reflection and the average expectation value of the superconducting gap is further strengthened by the analysis of $\kappa / \kappa_{N}$ spectroscopy for the $s_{ \pm}$ order parameter (Fig.~\ref{fig:fig5}). In our model Hamiltonian (which produces a Fermi surface shown in Fig.~\ref{fig:fig5}a), two superconducting gaps are observed, with a completely opened inner gap (Fig.~\ref{fig:fig5}b). Unlike Andreev measurements of d-wave superconductors, the Andreev current here persists at all energies and within both gaps for the $s_{ \pm}$ order parameter (Fig.~\ref{fig:fig5}b solid curves). As a result, the $\kappa$ spectrum is now very rich in detail and combines features from both s-wave and d-wave  (Fig.~\ref{fig:fig5}c). The persistence of Andreev current is consistent with our hypothesis regarding the expectation value of the superconducting gap. As seen in Fig.~\ref{fig:fig5}d, for the chosen parameters, the gap remains finite for any chemical potential. In general, it is of course possible that at some chemical potentials for the $s_{ \pm}$ or more complicated order parameters, the gap can become fully compensated, but unlike the d-wave, it is unlikely that complete compensation will be achieved for all chemical potentials. 

The shape of the $\kappa/ \kappa_N$ spectra can be accounted for by the competition between Andreev and quasiparticle tunneling. The conductance across the inner-most gap is almost exclusively due to Andreev reflection (Fig.~\ref{fig:fig5}b,e). Correspondingly, $\kappa / \kappa_{N}$ is at and above 2 inside the inner-most gap (Fig.~\ref{fig:fig5}c,f; the conductance falls below numerical accuracy for $|E| \lesssim \Delta$, where $\kappa$ cannot be resolved). At energies above both superconducting gaps, the Andreev conductance is much smaller than single electron tunneling, so that $\kappa / \kappa_{N}=1$, similar to the conventional s-wave order. However, at energies between the two superconducting gaps (between approximately 2.5 $\Delta$ and 3.5 $\Delta$ in Fig.~\ref{fig:fig5}b), both Andreev and quasiparticle tunneling can contribute to conductance, particularly with increased coupling. As a result, $\kappa / \kappa_{N}$ exhibits a transition from 1 to about 1.5 with increased coupling, and the exact shape of the transition in itself sensitively depends on the specific energy.

Finally, it is worth noting that the additivity of $\kappa$ enables rationalizing $\kappa$ spectroscopy at finite temperatures. At a phenomenological level, we can add a linear $\kappa$-term to Eq.~\ref{eq:kappaadd} weighted by its relative contribution, which will grow with increasing temperature (see Appendix~\ref{app:thermal} for details). If normalized by $\kappa_{N}$, this term will just be the conductance ratio of the quasiparticle current due to thermal excitations across the gap. This contribution will be dominant at smaller coupling, but will then compete with Andreev currents at stronger coupling due to the faster growth of Andreev contribution. Specifically, the persistence of Andreev reflection at stronger coupling enabled observing the $\kappa / \kappa_{N}$ trace of the superconducting gap in Pb up to the critical temperature \cite{Ko2022}.

\section{Conclusions}
We have elucidated the origin and variability of TAR spectroscopy, showing how
tunneling Andreev current depends on superconducting gap symmetry and
metal-superconductor coupling strength. Within the localized-junction model considered here, the likelihood of TAR is governed by
the matrix-element-weighted anomalous response. For momentum-independent tip-surface coupling, this response is proportional, to leading order, to the momentum-averaged superconducting gap. For the s-wave symmetry at zero
temperature, Andreev reflection solely determines mid-gap conductance; it is
entirely suppressed for d-wave order, and for $s_{\pm}$ symmetry, it remains
finite but competes with quasiparticle tunneling as a function of energy. The
relative decay rate ($\kappa/\kappa_N$) is additive, allowing all contributions
to be rationalized quantitatively in a parallel circuit framework. Notably, as the
normal-state conductance $G_N$ exceeds $0.1 G_0$, higher-order effects emerge in both
Andreev reflection and tunneling, markedly altering the $\kappa$-spectra and
providing signatures of the superconducting gap even for d-wave symmetry in the
absence of TAR. The $\kappa$-spectra therefore act as a fingerprint of
pairing symmetry complementary to traditional conductance-based spectroscopy.
Beyond our results,
further investigation can address the effects of a realistic band
structure, multi-channel junctions, topology, and magnetism. Ultimately, the
interplay of competing currents and the ability to quantify
$\kappa$-spectroscopy support its promise for identifying exotic quantum states
at the atomic scale.

\textbf{Acknowledgments}:
(PM) This work was supported, in part, at Clemson University by the State of South Carolina through funding for the Battelle Savannah River Alliance Workforce Development Program. Work at ORNL (BL) was supported by the U.S. Department of Energy, Office of Science, Basic Energy Sciences, Materials Sciences and Engineering Division. J.L.L.
acknowledges the computational resources provided by the Aalto Science-IT project and the financial support from  InstituteQ, the Finnish Quantum Flagship, the Research Council of Finland (no. 370912), the Finnish Centre of Excellence in Quantum Materials QMAT (no. 374166), and
the ERC Consolidator Grant ULTRATWISTROICS (Grant agreement no. 101170477).

\appendix
\setcounter{secnumdepth}{2}  
\setcounter{equation}{0}
\renewcommand{\theequation}{S\arabic{equation}}

\section{BTK expressions for conductance of a tunnel junction to s-wave superconductor analyzed as a power series}
\label{app:btk}

To clarify the meaning of $\kappa/\kappa_{N} > 2$ from an analytical perspective, we turn to the closed-form BTK expression for the energy-dependent junction conductance $G(E)$ of the s-wave superconducting junction in terms of gap energy $\Delta$ and coupling strength $\gamma$:

\begin{align}
G(E < \Delta) &= \frac{8\,\Delta^{2}\,\gamma^{4}}{\Delta^{2}(1+\gamma^{4})^{2} - E^{2}(1-\gamma^{4})^{2}} \label{eq:S1} \\
G(E > \Delta) &= \frac{2\gamma^{2}(E+\Omega)}{\Omega(1+\gamma^{4}) + 2E\gamma^{2}} \label{eq:S2}
\end{align}
where $\Omega = \sqrt{E^{2} - \Delta^{2}}$. Expanding both expressions for small coupling $\gamma \ll 1$ we obtain:

\begin{widetext}
\begin{align}
G(E < \Delta) &\approx \frac{8\Delta^{2}}{(\Delta^{2}-E^{2})}\,\gamma^{4}
- \frac{16\Delta^{2}(\Delta^{2}+E^{2})}{(\Delta^{2}-E^{2})^{2}}\,\gamma^{8}
+ \frac{8\Delta^{2}(\Delta^{2}+3E^{2})(3\Delta^{2}+E^{2})}{(\Delta^{2}-E^{2})^{3}}\,\gamma^{12} - \ldots \label{eq:S3} \\
G(E > \Delta) &\approx \frac{2(E+\Omega)}{\Omega} \Bigl[ \gamma^{2} -
	\frac{2E}{\Omega}\gamma^{4} +
	\frac{(4E^{2}-\Omega^{2})}{\Omega^{2}}\gamma^{6} - \ldots \Bigr]
	\label{eq:S4}
\end{align}
\end{widetext}

The expressions in Eqs.~\ref{eq:S3} and \ref{eq:S4} can be interpreted as net
conductance of parallel conductors, each represented by a polynomial term. The
leading order term of the Andreev conductance (Eq.~\ref{eq:S3}) scales as the square of
that of the above-gap conductance (Eq.~\ref{eq:S4}), as expected, reflecting a minimum of two barrier crossing
events required for a single Andreev reflection (each barrier crossing
contributes a probability of $\gamma^{2}$). Moreover, the higher order terms
in Eq.~\ref{eq:S3} ($\gamma^{8}$, $\gamma^{12}$) also have a clear semiclassical interpretation,
representing four and six total barrier crossings (shown schematically in Fig.~\ref{fig:fig3}d). Figure~\ref{fig:S1} shows the mid-gap and above-gap conductance as
a function of coupling, and the fitting of the above-gap conductance in
Figure~\ref{fig:S1}(b) using Eq.~\ref{eq:S4}, as explained in the main text.

\section{The effect of thermal broadening on $\kappa$ -- spectroscopy}
\label{app:thermal}

Arguably the most influence will be produced by thermal broadening, since other factors can be at least partially mitigated by optimization of the equipment and analysis. To illustrate thermal effect, we show below the evolution of the mid-gap $\kappa/\kappa_{N}$ from tunneling to near-contact with increasing temperature calculated from BTK model. Thermal broadening was included through Fermi-Dirac convolution, which is approximate but sufficient for the present analysis. As expected, the thermally excited quasiparticle currents strongly reduce the measured SC gap even at temperatures as low as $0.5\,T_{c}$ (compare Figure~\ref{fig:S2}(a) at $0.1\,T_{c}$ and \ref{fig:S2}(b) at $0.5\,T_{c}$). Electron tunneling due to thermally excited quasiparticle then becomes a parallel conductance channel to Andreev reflection at energies below superconducting gap, which is directly reflected by lowering of the $\kappa/\kappa_{N}$ below 2 at finite temperatures and small coupling (Figure~\ref{fig:S2}(c), $(G(4\Delta) < 10^{-2}G_{0})$). In fact, by the time the temperature reaches $\sim 0.2\,T_{c}$ tunneling Andreev reflection will be hardly detectable in good tunneling junctions. By contrast, detection of tunneling Andreev current is more robust at strong coupling ($G(4\Delta) > 10^{-1}G_{0}$) due to higher-order processes which contribute a $\kappa/\kappa_{N} > 2$ at 0K. Owing to these higher values, the additional steeper increase of conductance can still be detected despite the strong competition of electron tunneling, and BTK predicts that traces of Andreev current can be detected with $\kappa/\kappa_{N} > 1$ in the near-contact region up to at least $0.75\,T_{c}$, providing an independent verification of superconductivity particularly for strongly broadened, V-shaped superconducting gaps. At and above $T_{c}$, $\kappa/\kappa_{N} = 1$ due to suppression of superconductivity.

\bibliography{biblio}
\end{document}